\documentclass[useAMS,usenatbib]{mn2e}

\usepackage{amssymb}

\title[Shadowing unstable orbits]{Shadowing
  unstable orbits of the Sitnikov elliptic 3-body problem}
\author[D. J. Urminsky]{D. J. Urminsky\thanks{E-mail:
urminsky@astro.rit.edu}
\\
Department of Physics and Centre for Computational Relativity and
Gravitation, Rochester Institute of Technology, \\
85 Lomb Memorial Drive, Rochester, NY, 14623, USA}
\usepackage{amsfonts}
 \usepackage{graphicx}
\begin{document}


\pagerange{\pageref{firstpage}--\pageref{lastpage}} \pubyear{2002}

\maketitle

\label{firstpage}

\begin{abstract}
Errors in numerical simulations of gravitating  systems can be
magnified exponentially over short periods of time.  Numerical
shadowing provides a way of
demonstrating that the dynamics represented by numerical simulations are
representative of true dynamics.  Using the Sitnikov Problem as an
example, it is  demonstrated that unstable orbits of the 3-body problem can
be shadowed for long periods of time.  In addition, it is shown that
the stretching of phase space near escape and capture regions is a
cause for the failure of the shadowing refinement procedure.
\end{abstract}


\section{Introduction}

The sensitivity which $N$-body integrations exhibit to small changes
in initial conditions and to numerical errors has been an active area
of research since Miller's landmark study.  \cite{Miller} demonstrated the exponential
divergence of near-by orbits for systems with $N\leq 32$ and found
that the separation of nearby orbits increases rapidly when
close binary interactions occur.  He suggested that the
divergence of near-by  orbits is too rapid to be solely accounted by
binary interactions and suggests that there must be a collective
effect to account for the results.  However, \cite{Standish} showed
that the divergence rate was reduced if the potential was replaced
with a softened potential and concluded that the divergence is mainly
due to close binary interactions.

The dramatic effects of numerical errors on $N$-body
integrations was also demonstrated in an important paper by \cite{Lecar}.
After coordinating a study with 11 different integrations of the same
25-body problem for 2.5 crossing times, Lecar found that quantities
such as half mass radius and the moment of inertia  can
change by as much as 100 percent.
In a study with $N=3$, \cite{Hut} demonstrated that the amplification of
initial errors can increase by as much as $10^{20}$.  In addition,
they showed that the growth of errors during  close encounters
can be amplified by as much as $10^4$, however some of the growth can
be recovered  after the encounter is over.

The sensitivity to small changes in initial conditions and numerical
errors is a property associated with chaotic systems. A measure of the
sensitivity of numerical errors can be determined by the Lyapunov
exponent $\lambda$.  Early work suggested that the Lyapunov exponent
is inversely  proportional   to the crossing time $t_{cr}$ \citep{Kandrup,HeggieNbody,Goodman}. 
However, \cite{Goodman} suggest a dependence on $N$ of the form
$  \lambda^{-1} = t_{cr}/\log{N}$
or perhaps $\lambda^{-1} = t_{cr}/\log(\log(N))$, implying that as $N$
increases  the rate of separation decreases and the Lyapunov
exponent increases.  The $\log(N)$
dependence was later numerically verified by \cite{Merritt}.

Despite the difficulty calculating solutions to $N$-body integrations,
computers still remain a useful tool to study self gravitating
systems. If numerical errors in numerical solutions to the $N$-body
problem cause such drastic changes in the actual positions and
velocities of particles how can we trust the dynamics that these
solutions represent?
Shadowing is a way of proving that a true solution to a dynamical
system follows close to a numerical solution.  If  true orbits can be
found close to numerical orbits then the dynamics represented by the
numerical solutions represents true dynamics.

This study will discuss the existence of shadow orbits for the
gravitational 3-body problem. First,  definitions and concepts related to
shadowing of dynamical systems will be introduced.  Next, a
refinement procedure which makes corrections to  numerical orbits to
reduce the errors incurred at each time step will be presented.
The Sitnikov problem will then be presented and used as a simple
model to discuss  escape and capture of orbits.  An approximate
Poincar\'e map is then presented  to model orbits of the Sitnikov problem
and will be used in conjunction with the refinement procedure to
discuss the validity of numerical solutions by way of shadowing.
The failure of the refinement procedure to find shadow orbits will then be discussed and
regions of phase-space where the procedure fails will be delineated.
Finally, it will be demonstrated that the shadow times for this problem can be modeled
as a Poisson process.

\section{Shadowing}

Consider  the autonomous ordinary differential equation
\begin{equation}\label{eq:diffgen}
  \dot{\mathbf{x}} = f(\mathbf{x}),
\end{equation}
where $\mathbf{x}\in\mathbb{R}^n$ and $f:\mathbb{R}^n \rightarrow \mathbb{R}^n$ is a $C^1$ vector
field with the associated flow represented by $\vartheta^t$.  A sequence
of points $\{\mathbf{y}_k\}_{k=0}^M$ is said to be a
{\bf pseudo-orbit} if there is an associated bounded sequence
$\{h_k\}_{k=0}^M$ of positive time such that,
\begin{equation}
  | \mathbf{y}_{k+1} - \vartheta^{h_k}(\mathbf{y}_k)| < \delta,
\end{equation}
for $k=0,1,...,M$, where $\delta > 0$.  An example of a pseudo-orbit
is a numerical solution to (\ref{eq:diffgen}).  To show that a
pseudo-orbit represents some true dynamics for (\ref{eq:diffgen}), it
would be enough to show that a true orbit follows close to the
pseudo-orbit.  The pseudo-orbit described above is said to be 
{\bf shadowed} by a true orbit if there is a sequence of points
$\{\mathbf{x}_k\}_{k=0}^M$ and positive times $\{t_k\}_{k=0}^M$ with
$\vartheta^{t_k}(\mathbf{x}_k) = \mathbf{x}_{k+1}$ such that
\begin{equation}
  | \mathbf{x}_k - \mathbf{y}_k | < \epsilon, \;\; 
\end{equation}
and
\begin{equation}
 | t_k - h_k|<\epsilon,
\end{equation}
for $k=0,1,..,M$ and small $\epsilon>0$.  The sequence
$\{\mathbf{x}_k\}_{k=0}^M$ is known as a {\bf shadow-orbit}.  The
shadow-orbit is a true solution to (\ref{eq:diffgen}).

The first general contributions made on shadowing for dynamical systems
were the shadowing theorems of \cite{Anosov} and \cite{Bowen}.  Anosov
and Bowen considered hyperbolic systems and showed that  any
pseudo-orbit on a hyperbolic invariant set has a shadow-orbit.  These
theorems were generalized for pseudo-orbits in the vicinity of a
hyperbolic set \citep{Kato, Nadzieja,CoomesB}.   For non-hyperbolic
systems or for orbits which are far from hyperbolic invariant sets
these theorems do not apply.  Shadowing theorems do exist for
pseudo-orbits of non-hyperbolic systems and usually rely on
numerical verification of a theorem
\citep{CoomesA,ChowA,ChowPalmer,ChowVleck,Vleck}.

\subsection{Refinement procedure}\label{sec:GHYS}

Shadowing $N$-body simulations was first demonstrated by \cite{QT} and
\cite{HayesThesis}.  Both these studies considered the refinement
procedure found in \cite{GHYS} to find numerical shadows for the
$N$-body problem.
The refinement procedure  is a noise reduction
technique which can be used to show the existence of shadow-orbits.
This procedure will be presented for two dimensional dynamical maps,
however the procedure can easily be adapted for flows and has been extended to higher
dimensional systems by \cite{QT}.  

Consider the pseudo-orbit $\{\mathbf{p}_k\}_{k=0}^M$ of a map
$\mathbf{f}\in\mathbb{R}^2$.  The goal is to find a new less noisy orbit $\{
\hat{\mathbf{p}}_k \}_{k=0}^M$  close to the original
orbit.  Let $\mathbf{e}_k$ represent the one step error where
\begin{equation}\label{eq:ref1}
  \mathbf{e}_k = \mathbf{p}_k - \mathbf{f}(\mathbf{p}_{k-1}).
\end{equation}
The refined orbit is constructed by 
\begin{equation}\label{eq:ref2}
 \hat{\mathbf{p}}_k = \mathbf{p}_k + \mathbf{\Phi}_k,
\end{equation}
where $\bmath{\Phi}_k$ is the correction at time step $k$.  Combine
equations (\ref{eq:ref1}) and (\ref{eq:ref2}) to obtain
\begin{equation}\label{eq:ref3}
  \mathbf{\Phi}_k = \mathbf{f}(\hat{\mathbf{p}}_{k-1}) -
  \mathbf{e}_{k} - \mathbf{f}(\mathbf{\mathbf{p}_{k-1}}),
\end{equation}
where  $\hat{\mathbf{p}}_k =
\mathbf{f}(\hat{\mathbf{p}}_{k-1})$. Assuming that the correction,
$\mathbf{\Phi}_k$, is small, expand
$\mathbf{f}(\hat{\mathbf{p}}_{k-1})$ about $\mathbf{p}_{k-1}$ in a Taylor
series to get,
\begin{equation}\label{eq:ref4}
  \mathbf{f}(\hat{\mathbf{p}}_{k-1}) \approx
  \mathbf{f}(\mathbf{p}_{k-1}) + \bmath{L}_{k-1}\mathbf{\Phi}_{k-1},
\end{equation}
where $\bmath{L}_k$ is the linearized map at the $k$th time step.   
Substitute (\ref{eq:ref4}) into (\ref{eq:ref3}) to obtain
\begin{equation}\label{eq:ref5}
  \mathbf{\Phi}_{k} \approx \bmath{L}_{k-1}\mathbf{\Phi}_{k-1} -
  \mathbf{e}_k.
\end{equation}
It is also assumed that the linearized map has an expanding direction,
$\mathbf{u}_k$, and a contracting direction, $\mathbf{s}_k$, at each
time step $k$. With this assumption, the objective is to find the sequences
$\{\mathbf{\Phi}_k \}_{k=0}^M$ and $\{\mathbf{e}_k \}_{k=0}^M$  in the
coordinates $\{\mathbf{u}_k\}_{k=0}^M$ and $\{\mathbf{e}_k\}_{k=0}^M$ by
\begin{equation}\label{eq:us1}
    \mathbf{\Phi}_k  =  \alpha_k \mathbf{u}_k + \beta_k \mathbf{s}_k
\end{equation}
and
\begin{equation}\label{eq:us2}
  \mathbf{e}_k  = \eta_k \mathbf{u}_k + \zeta_k \mathbf{s}_k.
\end{equation}
The  expanding and contracting directions follow the linearized
maps,
\begin{equation}\label{eq:u}
\mathbf{u}_{k+1} = \bmath{L}_k\mathbf{u}_k,
\end{equation}
and
\begin{equation}\label{eq:s}
\mathbf{s}_{k+1} = \bmath{L}_k\mathbf{s}_k.
\end{equation}
For a random $|\mathbf{u}_0|=1$, equation (\ref{eq:u}) gives $\mathbf{u}_k$ aligned
with unstable direction at $\mathbf{p}_k$ after just a few iterations.
Starting with a random $\mathbf{s}_M$ and iterating (\ref{eq:s})
backwards gives $\mathbf{s}_k$ aligned with the stable direction at $\mathbf{p}_k$ after a few iterations.
  Substitute
(\ref{eq:us1}) and (\ref{eq:us2}) into (\ref{eq:ref5}) to get,
\begin{equation}\label{eq:ref6}
  \begin{array}{l}
    \alpha_{k+1} \mathbf{u}_{k+1} + \beta_{k+1} \mathbf{s}_{k+1}  =  \\
     \;\;\;\;\;\;\;\;\; \bmath{L}_k (\alpha_k \mathbf{u}_k + \beta_k
     \mathbf{s}_k) + (\eta_{k+1} \mathbf{u}_{k+1} + \zeta_{k+1}
     \mathbf{s}_{k+1}).
  \end{array}
\end{equation}
Substituting (\ref{eq:u}) and (\ref{eq:s}) into (\ref{eq:ref6}) yields
recursive relationships for $\{\alpha_k \}_{k=0}^N$ and $\{\beta_k
\}_{k=0}^N$ where,
\begin{equation}\label{eq:ab}
  \begin{array}{l}
  \alpha_{k+1} = |\bmath{L}_k\mathbf{u}_k|\alpha_k - \eta_{k+1}, \\
  \beta_{k+1} = |\bmath{L}_k \mathbf{s}_k|\beta_k - \zeta_{k+1}.
\end{array}
\end{equation}
Equations (\ref{eq:ab}) are made computationally stable by calculating
the coefficients $\alpha_k$  starting with $\alpha_M$ and iterating
backwards and the coefficients $\beta_k$ are calculated by
choosing an initial $\beta_0$ and iterating forwards.  The choice of
$\alpha_M$ and $\beta_0$ are arbitrary and  are
taken to be $\alpha_M = \beta_0 = 0$. Thus the sequence of  correction
coefficients are given by
\begin{equation}\label{eq:corrections}
  \begin{array}{ll}
    \alpha_M = 0, &  \alpha_k = \left( \alpha_{k+1} + \eta_{k_1}
    \right)/|\bmath{L}_k\mathbf{u}_k|, \\
    \beta_0 = 0, & \beta_k = |\bmath{L}_k\mathbf{s}_k|\beta_k - \zeta_{k+1},
  \end{array}
\end{equation} 
 where the values of $\eta_k$ and
$\zeta_k$ can be determined directly  from (\ref{eq:ref1}) and (\ref{eq:us2}).

Once $\{\hat{\mathbf{p}}_k\}_{k=0}^M$ has been found, the refinement
procedure can be iterated. Generally, the number of significant digits
doubles on each iteration of the process.  However, cases have been found
where the convergence is much slower or does not
converge.

The convergence of the refinement procedure does not in itself show
the existence of a shadow-orbit.  \cite{GHYS} provide a containment
procedure in two dimensions which rigorously proves the existence of a
shadow-orbit.  The containment
technique was later extended to three dimensional systems by
\cite{HayesThesis}.  A more practical approach for higher dimensional
systems was developed by \cite{SauerYorke}. They showed that for a given
pseudo-orbit, if the refinement procedure converges - to machine
precision - and certain quantities of a theorem remain bounded, then
the pseudo-orbit has a shadow-orbit.  

It has been found \citep{QT,Hayes}, that one can tell from
the convergence of the refinement procedure alone whether a given
pseudo-orbit can be shadowed.  So, if iterations of the
refinement procedure converge to a new orbit where the one-step
errors are the size of machine precision, then it is inferred that a
shadow-orbit exists for the given pseudo-orbit.  The new orbit
found by the refinement procedure is called a {\bf numerical shadow}.

\section{The Sitnikov Problem}

\begin{figure}
  \includegraphics[width=84mm]{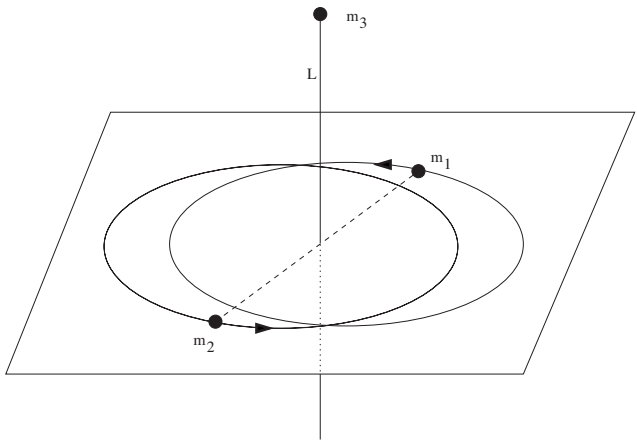}
  \caption{The Sitnikov Problem}
  \label{fig:Sitnikov}
\end{figure}

In this study of shadowing for the 3-body problem,  a
special configuration of the restricted 3-body problem known as the
Sitnikov problem will be considered.  The Sitnikov
problem is the problem of the motion of a mass-less
particle, $m_3$, on the axis of symmetry of an equal-mass binary (Figure
\ref{fig:Sitnikov}). Following \cite{Moser},  units are chosen such that
the gravitational constant $G=1$ and the total mass $M=1$.   Under
these conditions, the equation of motion for $m_3$ is given by
\begin{equation} \label{eqn:sitnikov}
\ddot{z} = -\frac{z}{\sqrt{z^2+r^2}^3},
\end{equation}
where $z$ is the position of $m_3$ and $r$ the distance from the centre of mass
to one of the binary masses. The distance $r$ can be
 approximated  to first order in the
eccentricity, $e$, by
\begin{equation}\label{eq:rapprox}
r\approx\frac{1}{2}(1-e\cos t),
\end{equation}
and  the specific energy of $m_3$ can be defined by
\begin{equation}\label{eq:spcenergy}
  E = \frac{1}{2}|\dot{z}|^2 - \frac{1}{\sqrt{r^2+z^2}}.
\end{equation}

\begin{figure*}
  \includegraphics[width=\linewidth]{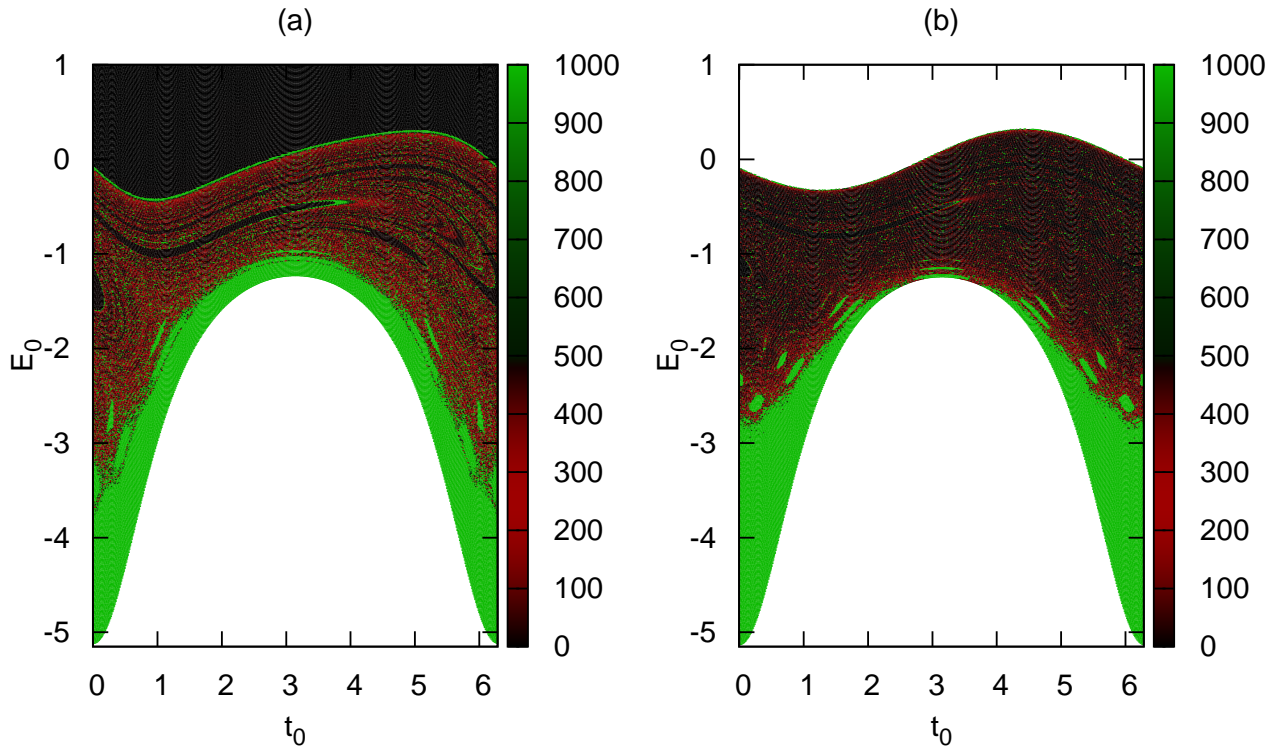}
  \caption{(a) Domain in $(t_0,E_0)$-space for the map $\phi$ where $e=0.61$.  Each
    point represents an initial condition and the associated colour
    represents the number of periods of the binary before escape.
    The green regions towards the bottom of the graph represent
    quasi-periodic orbits which remain bound for all time.  The solid
    black region at the top of the image are initial conditions
    outside of the domain of $\phi$.  The escape criterion used is
    effective in determining escape but crude in approximating the
    escape boundary on the SOS. (b)
    Domain in $(t_0,E_0)$-space for the map $\varphi$ where $e=0.61$.  The colour
    associated with each initial condition represents the number of
    periods of the binary before escape. In (b),  the number of
    periods of the binary  is
    determined by $t_M/2\pi$ where $M$ is the number of iterations of
    the map. The green regions
    represent quasi-periodic orbits which do not escape.}
  \label{fig:two}
\end{figure*}

\begin{figure}
  \includegraphics[width=\linewidth]{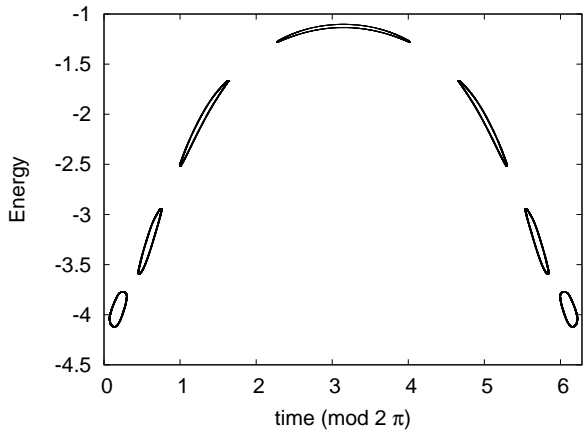}
  \caption{An example of a quasi-periodic orbit near a period 7 orbit for equation
    (\ref{eqn:sitnikov}) on the SOS $z=0$. Initial conditions are $\dot{z}(0) = 1.3$,
    $z(0) = 0.0$, $e=0.61$ and the phase of the binary is $.45$
    radians from pericentre.}
  \label{fig:new2}
\end{figure}

Taking the plane of motion of the binary ($z=0$) as a Surface Of
Section (SOS), consider a map, $\phi:(v_0,t_0)\rightarrow(v_1,t_1)$,
which takes $m_3$ from one crossing of the SOS to the next.  If
$m_3$ is on the SOS at time $t_0$, $\phi$ is a map which  brings
$v_0=\dot{z}(t_0)$ to time  $t_1>t_0$  where  $v_1=\dot{z}(t_1)$ and
$z(t_1)=0$. The map $\phi$ has an open domain $D_0$ in which every
point returns to the SOS. As time enters into the problem with period
$2\pi$, $D_0$ can be considered in
 polar co-ordinates where the radial
variable is $v$ and the angular variable is given by $t$.  
Alternatively, the domain $D_0$ can be considered on the surface of a
cylinder where
the initial position on the cylinder is defined by $t_0$ and $E_0$.
Figure \ref{fig:two} (a) shows the domain for $\phi$ in cylindrical coordinates.  The colour of each point
represents the number of periods of the binary before escape happens.
The green regions represent islands of quasi-periodic motion.  In
Figure \ref{fig:new2} an example of a quasi-periodic orbit which
visits the islands of stability in the vicinity of  a period 7 orbit
is provided.

\subsection{An approximate Poincar\'e map}

\cite{UrminskyHeggie} demonstrated that the Poincar\'e map $\phi$ with
 (\ref{eq:rapprox}) can be approximated by a simplectic map
$\varphi:(t_0,E_0)\rightarrow(t_1,E_1)$ where
\begin{equation}\label{eq:map}
\begin{array}{lcl}
  E_{1/2} & = & E_0 + a \cos(t_0) + b\sin(t_0), \\
  t_{1/1} & = & t_0 + 2 C (-E_{1/2})^{-3/2}, \\
  t_1 & = & t_{1/2} + 2 C (-E_{1/2})^{-3/2}, \\
  E_1 & = & E_{1/2} - a \cos(t_1) + b \sin(t_1),
\end{array}
\end{equation}
and $a$, $b$ and $C$ are constants.  The quantities $t_{1/2}$ and
$E_{1/2}$ are approximations of the time and energy values of $m_3$ at a
local maximum distance from the SOS. It is clear (Figure \ref{fig:two}
(a)) that the change in energy of $m_3$ from one crossing to the next
is periodic in time and the trigonometric terms in (\ref{eq:map}) can be
though of as the lowest order in a  Fourier approximation to this
change. The change in time is obtained by approximating the motion of
$m_3$ as Keplerian.  The constants $a$ and $b$ are proportional to the
eccentricity of the binary whose values can be shown to be,
\begin{equation}
  \begin{array}{lcl}
    a & \approx & 0.149 e \\
    b & \approx & 0.5075 e
  \end{array}
\end{equation}
and the constant $C = \pi/(2\sqrt{2})$.

\subsection{Escape and Capture}

Through interactions with the binary as it crosses the SOS, $m_3$ can gain
sufficient energy such that it leaves the SOS and does not return.  It
can be shown that for some positive time $t^*$ and positive
$\nu=(1-e)/2$, if
\begin{equation}\label{eq:cond1}
  \frac{1}{2}\dot{z}(t^*)^2 - \frac{1}{z(t^*)^2+\nu} > 0,
\end{equation} 
then  $|z(t)|\rightarrow 0$ as $t\rightarrow \infty$. Setting $z=0$ in
(\ref{eq:cond1}) gives a lower bound on the velocity of orbits which
escape on the SOS.  The solid black region at the top of Figure \ref{fig:two} (a)
demonstrates how this condition over estimates the escape
boundary.  All energy and time values in this region do not return to
the SOS.

\begin{figure}
  \includegraphics[width=\linewidth]{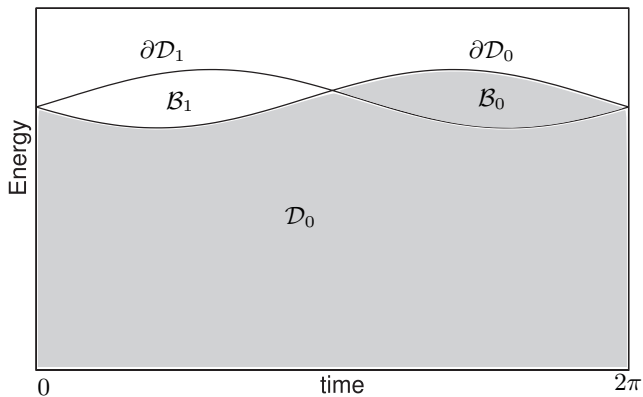}
  \caption{The curve $\partial \mathcal{D}_0$ represents a lower bound of
    energy and time values for which $\varphi$ is undefined.
    Similarly, the curve $\partial \mathcal{D}_1$ represents a lower bound of
    energy and time values for which the inverse map $\varphi^{-1}$ is undefined.
 The shaded region
    is the domain $\mathcal{D}_0$ for the map $\varphi$.  The two
    regions labeled $\mathcal{B}_0$ and $\mathcal{B}_1$ bounded by the
  curves $\partial \mathcal{D}_0$ and $\partial \mathcal{D}_1$ are the
  capture and escape regions respectively.}
  \label{fig:boundaries}
\end{figure}

The map, $\varphi$, provides an accurate way of determining escape and capture.
From equation (\ref{eq:map}) it is found that the mapping $\varphi$ is
defined in a region,
\begin{equation}\label{eq:Ebound}
 E_0 < -a\cos(t_0) - b\sin(t_0):=\partial \mathcal{D}_0,
\end{equation}
for
\begin{equation}\label{eq:Tbound}
  t_0 \in [0,2\pi] 
\end{equation}
as time enters into the mapping with period $2\pi$.
The curve $\partial \mathcal{D}_0$ is the escape boundary.  Time and energy values
above $\partial \mathcal{D}_0$ are said to have escaped.  
The domain, $\mathcal{D}_0$, can be defined by
 (\ref{eq:Ebound})
and (\ref{eq:Tbound}). 
  Initial conditions in $\mathcal{D}_0$ are mapped into the region, $\mathcal{D}_1$,
defined by
\begin{equation}\label{eq:Einbound}
  E < -a\cos(t) + b\sin(t):=\partial \mathcal{D}_1,
\end{equation}
for $t\in[0,2\pi]$.  Figure \ref{fig:boundaries} shows how the
boundaries $\partial \mathcal{D}_0$ and $\partial \mathcal{D}_1$ intersect.
 Orbits are mapped from the region under 
the curve $\partial \mathcal{D}_0$ to the region under the curve
$\partial \mathcal{D}_1$.
  The region $\mathcal{B}_0=\mathcal{D}_0\backslash\mathcal{D}_1$ represents energy
and time values for which orbits are captured.  In the context of the
differential equation, these are orbits which come from infinity and
get captured by the binary.  Similarly, the initial conditions in the
region $\mathcal{B}_1=\mathcal{D}_1\backslash\mathcal{D}_0$ are
energy and time values for which $\varphi$ is undefined.  Again, in
the context of the differential equation, the region $\mathcal{B}_1$
represents orbits which escape from the system.   Finally, note that initial
conditions for the differential equation are such that $\dot{z}>0$ and
$z=0$ on
the SOS. So from (\ref{eq:spcenergy}),  the initial energy can
be bounded from below by,
\begin{equation}\label{eq:below}
E > -1/|r(t_0)|,
\end{equation}
for $t_0 \in [0,2\pi]$.

\begin{figure}
  \includegraphics[width=\linewidth]{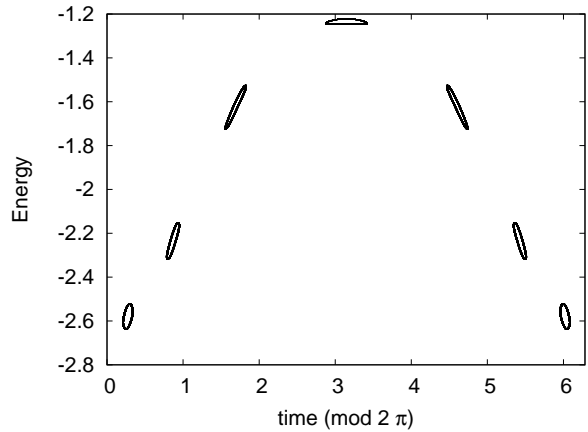}
  \caption{An example of a quasi-periodic orbit near a period 7 orbit
    for the map $\varphi$. Initial conditions are $t_0 = 6.01822$ and
    $E_0 = -2.5297$ for $e=0.61$.}
  \label{fig:new}
\end{figure}

Initial conditions are chosen in $\mathcal{D}_0$ with (\ref{eq:below})
for $e=0.61$ and plotted in Figure \ref{fig:two} (b).  The colour
of each point represents the number of periods of the binary
determined by $t_M/2\pi$ where $M$ is the number of iterations of the
map $\varphi$.  The green
 regions represent
stable motion whose orbits remain bounded.  As energy increases orbits
become unstable and escape from the system.
 Notice the similarities between Figure \ref{fig:two} (a) and
 \ref{fig:two} (b).  Both domains have islands representing stable orbits as
 well as large regions representing unstable orbits.  
  In Figure \ref{fig:new}
 an example of a quasi-periodic orbit near a period 7 orbit is
 provided.  In addition to the
similarities between Figure \ref{fig:two} (a) and (b),
\cite{UrminskyThesis} demonstrates that the map $\varphi$ satisfies
Lemmas similar to Lemmas 1-5 in \cite{Moser} (pages 87-91) and that
$\varphi$, like $\phi$,
possesses a hyperbolic invariant set on which $\varphi$ is
topologically equivalent to the shift map. 

\section{Results}
 
The map $\varphi$ provides a simple way of studying shadowing for orbits
like those of the Sitnikov problem.
 The approximate map is used to avoid integrating between
successive crossings of the SOS thus obtaining a tremendous speed up in
calculations.  In addition, the one step error can more easily be
controlled.
  At each time step uniformly
distributed noise $|\bmath{\delta}_k|\leq\delta$ is added to generate the
pseudo-orbit.  The refinement procedure is then used to reduce the
noise level to machine precision. Since $\varphi$ is a 2-dimensional mapping,
 the refinement procedure can be directly applied as shown in section
\ref{sec:GHYS}.

\subsection{Long lived orbits}

Using the containment and refinement procedure, \cite{GHYS}
successfully demonstrated the existence of shadows for pseudo-orbits
of length $10^7$ or more.  To test the algorithm  the
refinement procedure is applied to long lived orbits of the map
$\varphi$.  As seen in
Figure \ref{fig:two} (b), there are regions of stable motion where
orbits remain bounded forever. The refinement
procedure is applied to these orbits and it is found that most can be shadowed for many
iterations.  Some of these are shown in Figure \ref{fig:shadowing}.

\begin{figure}
    \includegraphics[width=\linewidth]{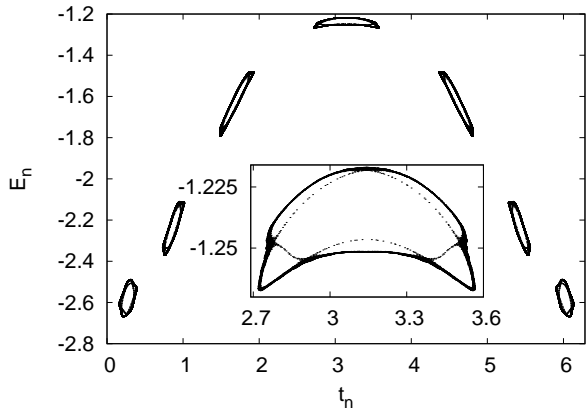}
  \caption{An example of a `sticky' orbit which remains close to
    islands of stable orbits for the map $\varphi$ for $500,0000$
    iterations.  The inset  box is
    a magnification of the upper most island of stability.}
  \label{fig:STICKY}
\end{figure}

\begin{figure}
    \includegraphics[width=\linewidth]{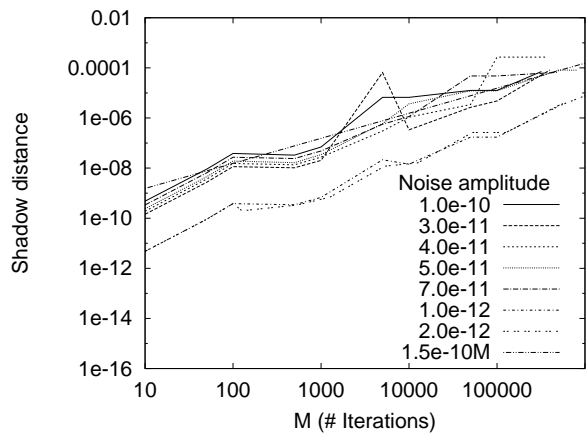}
  \caption{Number of iterations verses shadow distance $\epsilon$.}
  \label{fig:TvsEPSILON}
\end{figure}

As shown by \cite{Dvorak} for the Sitnikov problem, the map $\varphi$ has
`sticky' regions where orbits can be trapped for long periods of time
before escape. In Figure \ref{fig:STICKY} an example of a
sticky orbit trapped in the  vicinity of islands of stable
quasi-periodic orbits is shown.   The inset plot in Figure \ref{fig:STICKY} is a
magnification of the orbit near one of the islands.
By sampling the phase space around the islands  of stable motion, one
can find many sticky orbits which survive for long periods of time.
In Figure \ref{fig:TvsEPSILON}  the shadow distance is plotted against the
number of iterations of the map for several sticky orbits
where $e=0.61$.  It is shown that as the number of iterations increases,
the distance of the
numerical shadow  from the pseudo-orbit increases proportionally to
the number of iterations. 

\subsection{Shadowing capture orbits}

\begin{figure}
  \includegraphics[width=84mm]{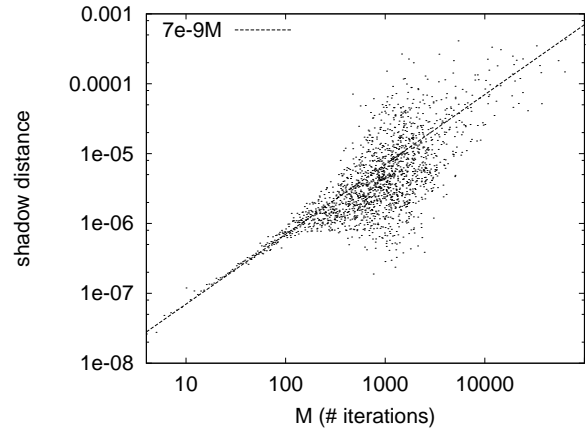}
  \caption{Each point is  average shadowing distance for the
    associate shadow length $M$ for  $10^5$ initial conditions in
    $\mathcal{D}_0$ where $e=0.25$. }
  \label{fig:eps09}
\end{figure}

Consider uniformly distributed  initial values in $\mathcal{B}_0$
(Figure \ref{fig:boundaries}) for $e=0.25$.  Initial values are
iterated forward for a maximum of 100000 iterations up to the penultimate
iteration before escaping. For each orbit, the number
of iterations, $M$, the orbit was `shadow-able' for as well as the
shadow-distance are recorded.  The orbits are binned into bins of length one
iteration and averaged over the bin.  The results are plotted in Figure
\ref{fig:eps09} where the dots represent the average shadow-distance
at each iteration of the map.  Note that as $M$ increases, there is
increasing variability on the distribution of average shadow distances.
The data can be fit with the curve $7\times 10^9 M$ which is similar to the results in
Figure \ref{fig:TvsEPSILON} where the shadow distance
is proportional to the orbit length.

\section{Where does shadowing fail?}\label{sec:5}

Numerical shadows have been found using the refinement
procedure for orbits whose length exceeds  $10^5$ iterations for the
map $\varphi$.  However, what happens when numerical shadows are not found?
What causes the refinement procedure to fail? First, it should be
noted that the failure of the refinement algorithm to
converge to a numerical shadow does not imply that there is not a
shadow-orbit for a given pseudo-orbit.  A shadow may still exist but
the refinement procedure was not able to converge towards it.  \cite{QT} and \cite{Hayes} found that
shadowing breaks down during close encounters between particles.  This
is due to the stretching of the velocity subspace during a close
encounter.  In the Sitnikov problem, $m_3$ interacts with the binary
on the SOS and the distance separating $m_3$ with the binary masses is
bounded from below (and above) on the SOS.  In contrast to the problems discussed in the above mentioned
studies  arbitrarily close
encounters do not occur in the Sitnikov problem.  However, escape and capture  occur during close encounters
with the binary  as $m_3$ crosses the SOS.  Near the escape and capture boundaries slight
changes in the energy of $m_3$ as it crosses the SOS can lead to
significant changes in the duration of successive crossings of the
SOS. The map provides a simple way of sampling the phase space
on the SOS to find regions where shadowing is more likely to fail. 

\begin{figure}
  \includegraphics[width=1.15\linewidth]{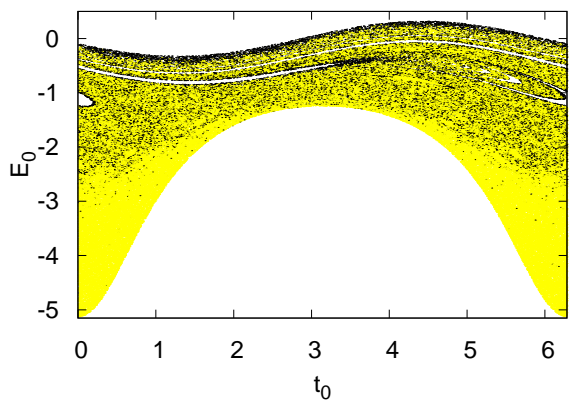}
  \caption{The figure shows one million initial conditions for the map
  $\varphi$ where $e=0.61$. The map $\varphi$ was applied to each
    initial condition  50,000 times or until the resulting orbit
    escaped. The colour associated with each point
    represents the successful application of the refinement
    algorithm.  Black represents initial conditions where the
    refinement procedure failed to converge.  Yellow represents the
    successful application of the refinement procedure. Only
    orbits which were longer than three iterations of $\varphi$ are
    considered. }
  \label{fig:shadowing}
\end{figure}

  Figure \ref{fig:shadowing} shows shadowing results of $10^6$
  initial conditions.  The colour of each
point represents either success, yellow, or failure, black, of the
refinement procedure. Note that only  orbits which survived more
than three iterations of the map are considered.  This is because the
choice of $\mathbf{u}_0$ and $\mathbf{s}_M$ would influence the results for short lived
orbits as (\ref{eq:u}) and (\ref{eq:s}) may not have had enough time
to align $\mathbf{u}_k$ and $\mathbf{s}_k$ in the proper directions.  From Figure
\ref{fig:shadowing} it can be seen that the 
 refinement procedure tends to fail near the escape boundary $\partial
 \mathcal{D}_0$. Note also that the refinement procedure fails
 near the boundaries of regions containing orbits which escape after three or less
 iterations.

The reason the refinement procedure fails in these regions is that
there is a stretching of subspace as orbits near the
boundary $\partial \mathcal{D}_0$.  At a given iteration $k$, the distance from
boundary, $\partial \mathcal{D}_0$, is given by,
\begin{equation}
  d = |E_k + a \cos(t_k) + b \sin(t_k)|.
\end{equation}
From the Jacobian of (\ref{eq:map}) it can
be shown that
\begin{equation}
 |\bmath{L}\mathbf{u}_k| \sim d^{-5/2}.
\end{equation} 
Thus, as $d\rightarrow 0$, the correction coefficients
$\alpha$ and $\beta$ go to 0 and $\infty$ respectively making it more
difficult for the refinement procedure to converge.

\begin{figure}
  \includegraphics[width=\linewidth]{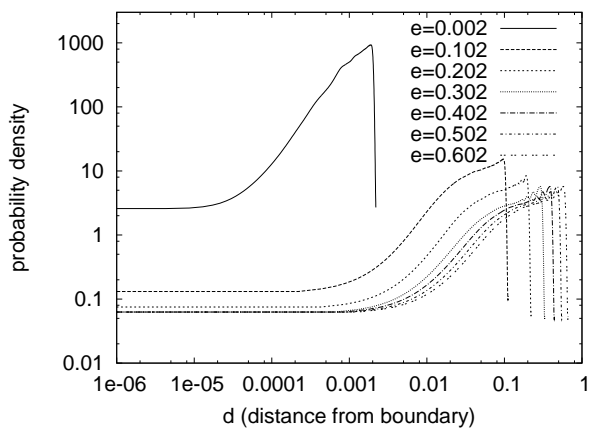}
  \caption{Probability density of close approaches to the escape
    boundary for shadow-able orbits.}
  \label{fig:Edist}
\end{figure}

Figure \ref{fig:Edist} shows the density of successfully shadowed
orbits based on the closest approach to the boundary $\partial \mathcal{D}_0$
for increasing eccentricity values.  For each shown eccentricity
value, we select 100000 uniformly distributed initial conditions in
the region defined by 
$t_0\in(\pi,2\pi)$ and $E_0\in(\partial \mathcal{D}_0 + 2 b\sin(t_0), \partial
\mathcal{D}_0)$. These boundaries describe a band of initial conditions bounded
above by the escape boundary. 
This band also encompasses the  capture region $\mathcal{B}_0$.
  The  drop in the density to the right of
each curve occurs at the distance between the lower boundary curve and
the escape boundary.  Note that the density drops off as initial
conditions approach the escape boundary.  Data was fitted using a
variable bandwidth kernel density function.

\subsection{Probability of capture}
 
It was found above that as orbits approach the escape boundary the
likelihood of an orbit being shadowed decreases.  This has an impact
on the shadow-ability of orbits in the capture region $\mathcal{B}_0$.
The capture region area is directly proportional to the eccentricity
of the binary.  As $e\rightarrow 0$, the initial conditions in
$\mathcal{B}_0$ become pushed up against the boundary
$\partial\mathcal{D}_0$.  It is expected then that for small
eccentricities, orbits would be less likely to be shadow-able.

\begin{figure}
    \includegraphics[width=84mm]{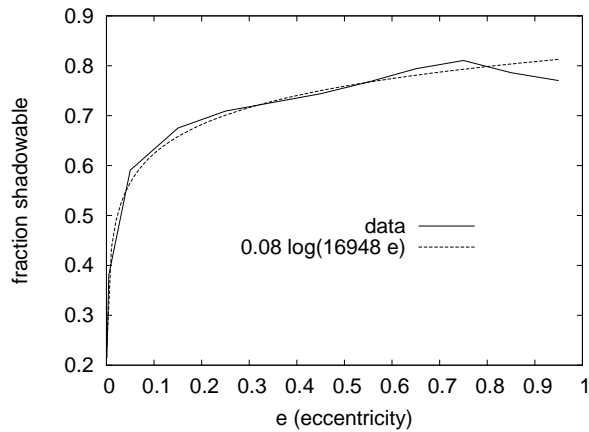}
  \caption{Fraction of capture orbits shadow-able using the refinement
  procedure for increasing eccentricities of the binary.}
  \label{fig:ECCvsPERCENT}
\end{figure}

To test this hypothesis,  $10^5$ uniformly distributed
initial conditions are selected in $\mathcal{B}_0$ and iterated forwards until
each orbit escapes.   This is performed for a variety of eccentricity values and
 the fraction of shadow-able orbits in each case is determined.  The results
are shown in Figure \ref{fig:ECCvsPERCENT}.
 The fraction of shadow-able orbits increases as the
eccentricity of the binary increases.   This is because the area of
the capture region increases proportionally to $e$.  As the area
increases, initial conditions can be selected at a much further
distance from the escape boundary making them more likely 
shadow-able. 

\subsection{Failure as a stochastic process}

\begin{figure}
  \includegraphics[width=\linewidth]{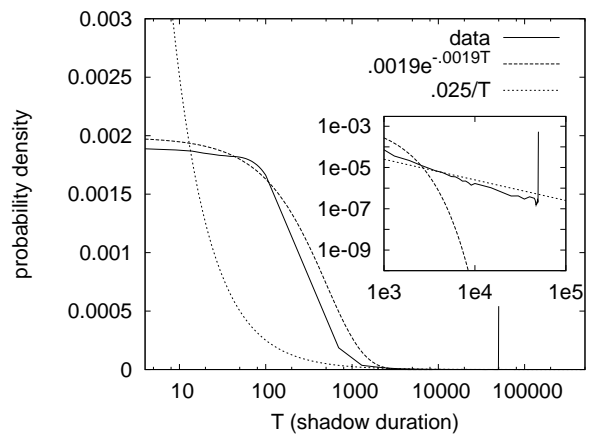}
  \caption{Probability density of shadow durations for the map
    $\varphi$ where e=0.61.  The amplitude of the one step noise was
    set at $10^{-9}$.  For shadow durations $T<5000$ the density can be
    approximated by an exponential distribution.  For larger $T$ the
    density is inversely proportional to $T$. }
  \label{fig:probden}
\end{figure}

\begin{figure}
  \includegraphics[width=\linewidth]{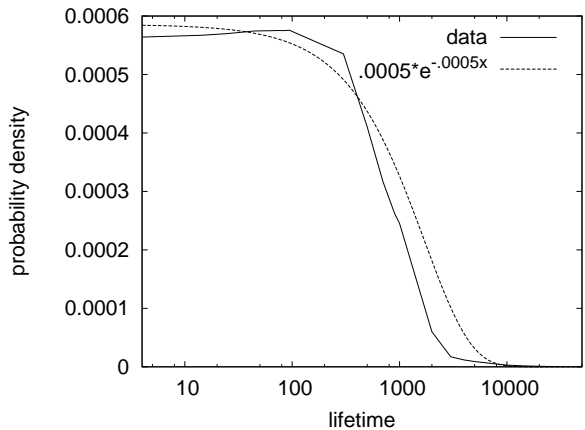}
  \caption{Probability density for the lifetime, $\sum_{k=0}^T t_k$, for the map
    $\varphi$ where e=0.61.  The amplitude of the noise is $10^{-9}$.  It
  was found that the distribution best fit  an exponential distribution.}
  \label{fig:probden2}
\end{figure}

  The failure of the refinement procedure can
happen at any point along the orbit and not necessarily at a close
approach to the escape boundary.   The shadow duration is defined 
as the number of iterations for which a given orbit can be shadowed.  For an
orbit $\{(t_i,E_i)\}_{i=0}^{M}$ the shadow duration, $T$, can take on
positive integer values $T<M$.

Consider  the initial conditions
for $e=0.61$ shown in Figure \ref{fig:shadowing}.  For each resulting
orbit, it is determined how long the orbit is shadow-able.
Figure \ref{fig:probden} provides some information on the distribution
of shadow lengths.  Initial conditions are chosen in $\mathcal{D}_0$ and iterated
forwards in time using (\ref{eq:map}).  Each orbit is iterated for
$50,000$ iterations or until the solution escapes.  
   The solid line in Figure \ref{fig:probden}
represents the density of the numerical experiments.  The spike at
50,000 iterations is mostly due to quasi-periodic orbits which remain
bounded for all time.   As shown in Figure \ref{fig:probden}, the
density can be approximated, for small iterations, by an exponential
density function given by $\xi \exp(-\xi x)$ for $\xi=0.0019$.  The
inset graph is a magnification of the density for $1000<M<50,000$.  In
this range, the density function is  better represented by the
function $.025/M$.

The map approximates the time between crossings of the SOS by
considering the motion of $m_3$ to be Keplerian.  Instead of
considering the distribution of the shadow duration in terms of the
number of iterations of $\varphi$ we can instead consider the
distribution of shadow times, $t_M$ where $M$ is the number of
iterations of the orbits for which it was shadow-able.  The solid line in
Figure \ref{fig:probden2} represents the probability density of shadow
time for the numerical experiments.  Again, the data can best be
approximated by the exponential density function for $\xi = .0005$.
The results found here are in agreement, for small shadow
durations, with previous results by
\cite{Hayes} which showed that shadow durations for larger $N$-body systems
have an exponential distribution and can be thought of as a Poisson process.

\section{Conclusions}

The above results confirm, for short lived orbits,
 previous
investigations \citep{Hayes} that showed numerical shadow durations,
$M$, for gravitating systems follow
a Poisson process with a exponential density function.  The result
found in this study suggests for
longer lived orbits, the density function is better approximated  by a
function proportional to $1/M$.  This may be because the population of
longer lived
orbits tends to be dominated by stable orbits, however this has not
been investigated.

In section \ref{sec:5}, areas of phase-space where the refinement
procedure is more likely to fail are characterized.
These areas are near
escape boundaries where  there is sufficient stretching of phase-space
to cause the refinement procedure to fail to converge to a less noisy
orbit.  Interestingly this seems to be due to the
growth of the variational equations over one time step. This does not
rule out the failure of the the refinement procedure by the accumulative
effect of the growth of the variational equation associated with large
Lyapunov exponents as discussed by \cite{Zhu}.

In Figure \ref{fig:ECCvsPERCENT} it is demonstrated that as the volume of
phase-space representing capture orbits decreases, it becomes increasingly
 difficult to shadow capture orbits.  This is a result of the
distribution of failures of the refinement procedure seen in Figure
\ref{fig:Edist}.   As the volume of  phase-space associated with
capture decreases, capture orbits get pushed up against the boundary
$\partial \mathcal{D}_0$ where the one-step growth of the variational
equations causes the refinement procedure to fail.

Finally, it was found that the shadow distance for an orbit is
proportional to the number of iterations of the map (Figures
\ref{fig:TvsEPSILON} and \ref{fig:eps09}).  It was noticed that if 
in addition to $t_1$ and $E_1$, orbits were required to be shadow-able
at the half steps $t_{1/2}$ and $E_{1/2}$, then initially shadow-able
orbits continued to be shadow-able.
  When shadowing at the half step was
required, the shadow distance typically increased by about a factor of
two.

The Sitnikov problem discussed in this study provides a straight
forward way of characterizing a domain of initial conditions as well
as regions of stable and unstable motion. Work in progress considers
slight
changes to the Sitnikov problem in order  to study shadowing of
unstable orbits.  For example, \cite{Soulis1} consider slight
perturbations to the mass and  position (away from the $z$-axis) of
$m_3$ and delineate regions of stable and unstable motion. It would be
expected that, like the results found in this study, shadowing with the
refinement procedure breaks down near boundaries of escape for
unstable orbits.   In fact, the break down of the refinement procedure
near escape boundaries would be expected for  general 3-body
configurations.  As solutions approach parabolic escape boundaries,
an orbit can undergo increasingly  long ejections from the left-over
binary system.  Small changes in the energy of an orbit in this region
can cause significant changes in the time of return for the orbit.
 If the refinement procedure could make changes to the
orbits so as to conserve the energy of the ejected body it might
improve the success rate of the refinement procedure.
 Finally, the Sitnikov 4-body problem \citep{Soulis2}
provides a starting point for examining the relationship between the
shadowing distance and the number of bodies.  Extra bodies can be
added in circular orbits about the center of mass. \cite{Hayes} demonstrates
that as the number of moving bodies in a fixed potential increases the shadow durations decrease.
It would be of interest to determine if a similar relationship holds
for the Sitnikov $N$-body problem.

It should be stressed again that the failure of the  refinement procedure
does not necessarily mean that a shadow does not exist for a given
pseudo-orbit.   It may very well be that shadows do exist for 
orbits in regions where the refinement procedure fails.  We are
encouraged that this may be the case.  Both the Sitnikov problem and the
approximate Poincar\'e map possess a hyperbolic invariant set, $\Lambda$, near the
escape boundaries (see \cite{Moser} and \cite{UrminskyThesis}
respectively).  Despite the fact that $\Lambda$ is near the boundary
$\partial \mathcal{D}_0$, the shadowing theorems by \cite{Anosov} and
\cite{Bowen} guarantee that any pseudo-orbit on $\Lambda$ has an
associated shadow-orbit.  This demonstrates that being in the vicinity on the escape
boundary does not necessarily rule out the existence of
shadow-orbits.

\section*{Acknowledgments}
DU was supported by the National Aeronautics and Space Administration
through grant NNX-07AH15G. The author would   like to thank
D. Heggie, D. Merritt and D. Dicken for their helpful
suggestions.  In addition, the author would like to thank the
anonymous referee for his/her careful reading of the manuscript and useful suggestions.

\label{lastpage} 

\end{document}